\documentclass[a4paper,12pt]{article}
\usepackage[english]{babel}
\usepackage{graphicx}
\usepackage{times}
\usepackage[dvips,unicode,colorlinks,linkcolor=blue,citecolor=blue,urlcolor=blue]{hyperref}
\begin{document}
\author{O. V. Gendelman$^{1,2}$ \and
        A. V. Savin$^{1,3}$}

\title{Heat Conduction in One-Dimensional chain of Hard Discs with Substrate Potential}

\maketitle
\footnotetext[1]{Institute of Chemical Physics RAS, Kosygin str. 4, Moscow 119991, Russia}
\footnotetext[2]{ovgend@center.chph.ras.ru} \footnotetext[3]{asavin@center.chph.ras.ru}

\tableofcontents
\begin{abstract}
Heat conduction of one-dimensional chain of equivalent rigid particles in the field of external
on-site potential is considered. Zero diameters of the particles correspond to exactly
integrable case with divergent heat conduction coefficient. By means of simple analytical model
it is demonstrated that for any nonzero particle size the integrability is violated and the
heat conduction coefficient converges. The result of the analytical computation is verified by
means of numerical simulation in a plausible diapason of parameters and good agreement is
observed
\end{abstract}

\section{Introduction}
Heat conductivity in one-dimensional (1D) lattices is a well known classical problem related to
the microscopic foundation of Fourier's law. The problem started from the famous work of Fermi,
Pasta, and Ulam (FPU) \cite{p1}, where an abnormal process of heat transfer was detected for
the first time. Non-integrability of a system is a necessary condition for normal heat
conductivity. As it was demonstrated recently for the FPU lattice \cite{p2,p3,p4}, disordered
harmonic chain \cite{p5,p6,p7}, diatomic 1D gas of colliding particles \cite{p8,p9,p10,p11},
and the diatomic Toda lattice \cite{p12}, non-integrability is not sufficient in order to get
normal heat conductivity. It leads to linear distribution of temperature along the chain for
small gradient, but the value of heat flux is proportional to $1/N^\alpha$, where $N$ is the
number of particles of the chain, and number exponent $0<\alpha<1$. Thus, the coefficient of
heat conductivity diverges in the thermodynamic limit $N\rightarrow\infty$. Analytical
estimations \cite{p4} have demonstrated that any chain possessing an acoustic phonon branch
should have infinite heat conductivity in the limit of low temperatures.

From the other side, there are some artificial systems with on-site potential (ding-a-ling and
related models) which have normal heat conductivity \cite{p13,p14}. The heat conductivity of
the Frenkel-Kontorova chain was first considered in Ref. \cite{p15}. Finite heat conductivity
for certain parameters was observed for Frenkel-Kontorova chain \cite{p16}, for the chain with
sinh-Gordon on-site potential \cite{p17}, and for the chain with $\phi^4$ on-site potential
\cite{p18,p19}. These models are not invariant with respect to translation and the momentum is
not conserved. It was supposed that the on-site potential is extremely significant for normal
heat conduction \cite{p18} and that the anharmonicity of the on-site potential is sufficient to
ensure the validity of Fourier's law \cite{p20}. A recent detailed review of the problem is
presented in Ref. \cite{p21}.

Probably the most interesting question related to heat conductivity of 1D models (which
actually inspired the first investigation of Fermi, Pasta and Ulam [1]) is whether small
perturbation of an integrable model will lead to convergent heat conduction coefficient. One
supposes that for the one-dimensional chains with conserved momentum the answer is negative
\cite{p21a}. Still, normal heat conduction has been observed in some special systems with
conserved momentum \cite{p22,p23,p24}, but it may be clearly demonstrated only well apart from
integrable limit.

The situation is not so clear in the systems with on-site potential. Although it was supposed
that non-integrable system without additional integrals of motion would have convergent heat
conductivity \cite{p21a}, no rigorous proof was presented. From the other side, recent attempt
of numerical simulation of heat transfer in Frenkel-Kontorova model \cite{p25} demonstrated
that because of computational difficulties no unambiguous conclusion can be drawn whether heat
conduction is convergent for all finite values of perturbation of the integrable limit system
(linear chain or continuous sine-Gordon system).

It seems that computational difficulties of investigation of heat conduction in a vicinity of
integrable limit are not just issue of weak computers or ineffective procedures. In the systems
with conserved momentum divergent heat conduction is fixed by power-like decrease of heat flux
autocorrelation function with power less than unity. Still, for the systems with on-site
potential exponential decrease is more typical \cite{p25}. For any fixed value of the exponent
the heat conduction converges; if the exponent tends to zero with the value of the perturbation
of the integrable case, then for any finite value of the perturbation the characteristic
correlation time and length will be finite but may become very large. Consequently, they will
exceed any available computation time or size of the system and still no conclusion on the
convergence of heat conduction will be possible.

In this respect it is reasonable to mention the findings of our papers \cite{p23,p24,p25}
concerning the transitions from infinite to finite heat conduction in the chain of rotators and
Frenkel - Kontorova system. These findings were criticized in paper \cite{p25a} by providing
the computation results for larger systems. We agree with the authors that the results of
papers \cite{p23,p24} do not prove the reality of the above transition. In fact, we have
pointed it out in \cite{p25}. From the other side, it is clear that such more extended
numerical simulations cannot prove that the transitions do not exist at all. Probably, such
conclusion may be driven if, with the help of some not yet developed method, the simulation
length or effective time will be extended by many orders of magnitude.

The other way to overcome this difficulty is to construct a model, which will be, at least to
some extent, tractable analytically and will allow one to predict some characteristic features
of the heat transfer process and the behavior of the heat conduction coefficient. Afterwards
the numerical simulation may be used to verify the as\-sumptions made in the analytic
treatment. To the best of our knowledge, no models besides pure harmonic chains were treated in
such a way to date. Introduction and investigation of a nontrivial model of this sort is a
scope of present paper.

We are going to demonstrate that there exist models which have integrable system as their
natural limit case, small perturbation of the integrability immediately leading to convergent
heat conduction. The mechanism of energy scattering in this kind of systems is universal for
any temperature and set of the model parameters. The simplest example of such model is
one-dimensional set of equal rigid particles with nonzero diameter $(d>0)$ subjected to
periodic on-site potential. This system is completely integrable only if $d=0$. It will be
demonstrated that any $d>0$ leads to effective mixing due to unequal exchange of energy between
the particles in each collision. This mixing leads to diffusive mechanism of the heat transport
and, subsequently, to convergent heat conduction.

\section{Description of the model}

\begin{figure}[t]
\begin{center}
\includegraphics[angle=0, width=0.75\textwidth]{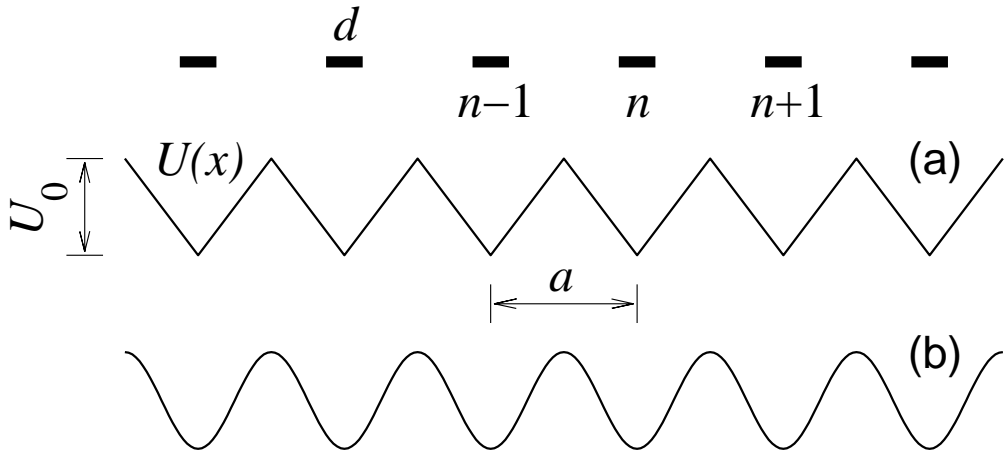}
\end{center}
\caption{\label{fig1}\protect\small
       Sketch of the hard-disk chain exposed to periodic on-site potential $U(x)$ ($a$ is the
       period of the potential, $U_0$ -- its height, $d$ is the diameter of the disks).
       Piecewise-linear potential (\ref{f10}) (a) and sinusoidal potential (\ref{f20}) (b)
       are plotted.
        }
\end{figure}

Let us consider the one-dimensional system of hard particles with equal masses subject to
periodic on-site potential. The Hamiltonian of this system will read
     \begin{equation}
     {\cal H}=\sum_n\{ \frac12 M\dot{x}_n^2+ V(x_{n+1}-x_n)+U(x_n)\},
     \label{f1}
     \end{equation}
where $M$ -- mass of the particle, $x_n$ -- coordinate of the center of the $n$-th particle,
$\dot{x}_n$ -- velocity of this particle, $U(x)$ -- periodic on-site potential with period $a$
$[U(x)\equiv U(x+a)]$. Interaction of absolutely hard particles is described by the following
potential
     \begin{equation}
     V(r) = \infty~\mbox{if} ~r \le d~~~\mbox{and}~~~
     V(r) = 0~\mbox{if}~ r> d,
     \label{f2}
\end{equation}
where $d$ is the diameter of the particle. This potential corresponds to pure elastic impact
with unit recovery coefficient. Sketch of the model considered is presented at Fig. \ref{fig1}.

It is well-known that the elastic collision of two equal particles with collinear velocity
vectors leads to exchange of their velocities. An external potential present does not change
this fact, since the collision takes zero time and thus the effect of the external force on the
energy and momentum conservation is absent.

The one-dimensional chain of equivalent hard particles without external potential is a paradigm
of the integrable nonlinear chain, since all interactions are reduced to exchange of
velocities. In other words, the individual values of velocities are preserved and just
transferred from particle to particle. It is natural therefore to introduce quasiparticles
associated with these individual values of velocities. They will be characterized by a pair of
parameters $(E_k, {\bf n}_k)$, where $E_k = v_k^2/2$ is an energy of the quasiparticle, ${\bf
n}_k$ is a unit vector in a direction of its motion. Every particle in every moment "carries"
one quasiparticle. The elastic collision between the particles leads to simple exchange of
parameters of the associated quasiparticles, therefore the quasiparticles themselves should be
considered as non-interacting.

\section{Analytical study}

The situation changes if the external on-site potential is present. It is easy to introduce
similar quasi\-particles ($E_k$ will be a sum of kinetic and potential energy). The unit vector
{\bf n} of each quasiparticle between subsequent interactions may be either constant (motion in
one direction) or periodically changing (vib\-ration of the particle in a potential well). In
every collision the particles exchange their velocity vectors, but do not change their
positions. Consequently two quasiparticles interact in a way described by the following
relationships:
        \begin{eqnarray}
        E'_1&=&E_1+U(x_c+d/2)-U(x_c-d/2), \nonumber \\
        E'_2&=&E_2-U(x_c+d/2)+U(x_c-d/2), \nonumber \\
        {\bf n}'_1&=&{\bf n}_1, \label{f3} \\
        {\bf n}'_2&=&{\bf n}_2. \nonumber
        \end{eqnarray}
The values denoted by apostrophe correspond to the state after the collision, $x_c$ is a point
of contact between the particles. It should be mentioned that in the case of nonzero diameter
the quasiparticles are associated with the centers of the carrying particles.

If the diameter of the particles is zero, then the additives to the energies in the first two
equations of system (\ref{f3}) compensate each other and the energies of the quasiparticles are
preserved in the collision. Therefore effectively the interaction between the quasiparticles
disappears and the chain of equal particles with zero size subject to any on-site potential
turns out to be completely integrable system. Thus, contrary to some previous statements it is
possible to construct an example of strongly nonlinear one-dimensional chain without momentum
conservation, which will have clearly divergent heat conductivity (even linear temperature
profile will not be formed).

The situation differs if the size of the particles is not zero, as the individual energies of
the quasiparticles are not preserved in the collisions. In order to consider the effect of such
interaction we propose simplified semi-phenomenological analytical model.

After $l$ collisions the energy of the quasiparticle will be
       \begin{equation}
       E(l)=E_0+\sum_{j=1}^l\Delta E_j,~~\Delta E_j=U(x_j+d/2)-U(x_j-d/2),
       \label{f4}
       \end{equation}
where $j$-th collision takes place in point $x_j$, $E_0$ is the initial energy of the
quasiparticle. Now we suppose that the coordinates of subsequent contact points $\{
...,x_{j-1}, x_j,x_{j+1},...\}$, taken by modulus of the period of the on-site potential, are
not correlated. Such proposition is equivalent to fast phase mixing in a system close to
integrable one and is well-known in various kinetic problems \cite{p26}. The consequences of
this proposition will be verified below by direct numerical simulation.

Average energy of the quasiparticle is equal to $\langle E_0 \rangle$ over the ensemble of the
quasiparticles, as obviously $\langle\Delta E_j \rangle=0$. Still, the second momentum will be
nonzero:
       \begin{equation}
       \langle(E(l) - E_0 )^2\rangle = l\langle(U(x+d/2)-U(x-d/2))^2\rangle_x
       \label{f5}
       \end{equation}
The right-hand side of this expression will depend only on the exact shape of the potential
function
       \begin{equation}
       \langle(E(l) - E_0 )^2\rangle = lF(d),~~
       F(d)=\frac1a\int_0^a[U(x+d/2)-U(x-d/2)]^2dx,
       \label{f6}
       \end{equation}
The last expression is correct only at the limit of high temperatures; it neglects the fact
that the quasi\-particle spends more time near the top of the potential barrier due to lower
velocity.

Let us consider the quasiparticle with initial energy $E_0> U_0$, where $U_0$ is the height of
the potential barrier. Therefore vector ${\bf n}$ is constant. Equations (\ref{f5}) and
(\ref{f6}) describe random walks of the energy of the quasiparticle along the energy scale
axis. Therefore after certain number of steps (collisions) the energy of the quasiparticle will
enter the zone below the potential barrier $E(l)< U_0$. In this case the behaviour of the
quasiparticle will change, as constant vector ${\bf n}$ will become oscillating, as described
above. After some additional collisions the energy will again exceed $U_0$, but the direction
of motion of the quasiparticle will be arbitrary. It means that the only mechanism of energy
transfer in the system under consideration is associated with diffusion of the quasiparticles,
which are trapped by the on-site potential and afterwards released in arbitrary direction. Such
traps-and-releases resemble Umklapp processes of phonon-phonon interaction \cite{p26}, but
occur in a purely classic system.

The diffusion of the quasiparticles in the chain is characterized by mean free path, which may
be evaluated as
       \begin{equation}
       \lambda \sim\frac{2a\langle(U_0-E_0)^2\rangle}{n_cF(d)}
       \sim\frac{2a[2(k_BT)^2-2U_0k_BT+U_0^2]}{n_cF(d)},
       \label{f7}
       \end{equation}
where $n_c$ is a number of particles over one period of the on-site potential (concentration).
Coefficient 2 appears due to equivalent probability of positive and negative energy shift in
any collision, $T$ -- tempera\-ture of the system, $k_B$ -- Boltzmann constant.

Average absolute velocity of the quasiparticle may be estimated as
       \begin{equation}
       \langle |v| \rangle\sim \frac{a}{{a - n_cd}}\sqrt{\frac{\pi k_BT}{2}}
       \label{f8}
       \end{equation}
Here the first multiplier takes into account the nonzero value of $d$ and absolute rigidity of
the particles. The second one is due to standard Maxwell distribution function for 1D case.

Heat capacity of the system over one particle is unity, as the number of degrees of freedom
(i.e. the number of quasiparticles) coincides with the number of the particles and does not
depend on the temperature and other parameters of the system. Therefore the coefficient of
heat conductivity may be estimated \cite{p26} as
       \begin{equation}
       \kappa\sim\lambda\langle |v|\rangle\sim\frac{2a^2}{n_c(a-n_cd)}
       \frac{2(k_BT)^2-2U_0k_BT+U_0^2}{F(d)}\sqrt{\pi k_B T/2}
       \label{f9}
       \end{equation}
It is already possible to conclude that according to (\ref{f9}) regardless the concrete shape
of potential $U(x)$ in the limit $d\rightarrow 0$ we have $F(d)\rightarrow 0$ and therefore
$\kappa\rightarrow\infty$, although for every nonzero value $d$ the heat conductivity will be
finite. Therefore unlikely known models with conserved momentum the small perturbation of the
integrable case $d=0$ immediately brings about convergent heat conductivity.

\section{Numerical simulations}

It is convenient to introduce the dimensionless variables for the following numerical
simulation. Let us set the mass of each particle $M=1$, on-site potential period $a=2$, its
height $U_0=1$, and Boltzmann constant $k_B=1$ in all above relationships. We suppose that the
chain contains one particle per each period of the potential, i.e. that $n_c=1$, and the
particle diameter $0<d<2$.

Let us consider periodic piecewise linear on-site potential
      \begin{eqnarray}
       U(x) &=& x~~\mbox{if}~~x\in [0,1], \nonumber \\
       U(x) &=& 2-x~~\mbox{if}~~x\in [1,2], \label{f10}\\
       U(x+2l)&=&U(x)~~\mbox{for}~~x\in [0,2],~~l=0,\pm1,\pm2,... \nonumber
      \end{eqnarray}
(the shape of the potential is presented at Fig. \ref{fig1}). Then it follows from (\ref{f9})
that the non-dimensional heat conduction coefficient is expressed as
      \begin{equation}
      \kappa=\frac{8(2T^2-2T+1)}{(2-d)F(d)}\sqrt{\pi T/2},
      \label{f11}
      \end{equation}
where function
      \begin{eqnarray}
      F(d) &=& d^2-2/3d^3,~~\mbox{for}~~0<d\le1 \label{f12} \\
      F(d) &=& -4/3+4d-3d^2+2/3d^3,~~\mbox{for}~~1\le d < 2. \nonumber
     \end{eqnarray}

The numerical scheme for solving the equations of motion describing the dynamics of the 1D
hard-point gas has been developed in a series of papers \cite{p27,p8,p28}.

Dynamics of the system of particles with potential of the nearest-neighbor interaction
(\ref{f2}) and piece\-wise linear on-site potential (\ref{f10}) may be described exactly.
Between the collisions each particle moves under constant force with sign dependent on the
position of the particle. Therefore the coordinate of each particle depends on time $t$ as
piecewise parabolic function which may be easily computed analytically. If the particle centers
are situated at distance equal to $d$, then elastic collision occurs. The particles exchange
their momenta as described above and afterwards the particle motion is again described by
piecewise parabolic functions until the next collision.

Let us consider finite chain of $N$ particles with periodic boundary conditions. Let at the
moment $t=0$ one particle be at each potential minimum and let us choose Boltzmann's
distribution of the initial velocity. Solving the equations of motion, we find a time $t_1$ of
the first collision between some pair of the adjacent particles, next a time $t_2$ of the
second collision, in general between another pair of the adjacent particles, and so on. As a
result, we obtain a sequence $\{t_i,n_i\}_{i=1}^\infty$, where $t_i$ is the time of the $i$th
collision in the system, and $n_i$ and $n_i+1$ are the particles participating in this
collision. Since we need to implement numerical simulations as long as possible, in order to
find the time asymptotic of the heat flux autocorrelation function entering the Green-Kubo
formula, we use the numerical scheme of paper \cite{p9}. First, we incorporate the energy
change of the $n_i$th particle during the $i$th collision  as
                 $$
                 \Delta E_{n_i}=\frac12({v'_{n_i}}^2-{v_{n_i}}^2)
                 =\frac12({v_{n_i+1}}^2-{v_{n_i}}^2).
                 $$
Next, we introduce a time step $\Delta t$, which is significantly less than the simulation
time, but satisfies the inequality $\Delta t\gg t_0$, where
$t_0=\lim_{i\rightarrow\infty}(t_i/i)$ is the mean time between successive collisions. Then,
for each $k=0,1,...$, we define the local energy flow as a piecewise constant (in time)
function
      \begin{equation}
      j_n(t)=\frac{a}{\Delta t}\sum_{i\in I_{kn}}\Delta E_{n_i},~~ k\Delta t\le t<(k+1)\Delta t,
      \label{f13}
      \end{equation}
where the integer sets $I_{kn}$'s are defined by
       $$
       I_{kn}=\{i|~k\Delta t\le t_i<(k+1)\Delta_t,~n_i=n\}.
       $$
The set $I_{kn}$ takes into account those collisions that occur between particles $n$ and $n+1$
during the time interval $k\Delta t\le t<(k+1)\Delta_t$. Equilibration times were typically
occurring in the system of the  order $10^6$. After these times have passed, we define the
time-averaged local thermal flow
       \begin{equation}
       J_n=\langle j_n(t)\rangle_t\equiv\lim_{t\rightarrow\infty}\frac1t\int_0^tj_n(\tau)d\tau
       \label{f14}
       \end{equation}
and the temperature distribution $T_n=\langle v_n^2(t)\rangle_t$, where $v_n(t)$ is the
velocity of particle $n$ calculated at a time $t$. To find these averaged quantities, we have
used times up to $10^7$.

To find the heat flux autocorrelation function $C(t)$ numerically, we calculated the time
mean\linebreak $\langle J(\tau)J(\tau-t)\rangle_\tau/NT^2$, with $J(t)=\sum_nj_n(t)$ being the
total heat flow through the gas/chain system consisting of $N=500$ particles and temperature
$T=\sum_nT_n/N$ averaged over $10^4$ realizations of initial thermalization.

\begin{figure}[t]
\begin{center}
\includegraphics[angle=0, width=0.85\textwidth]{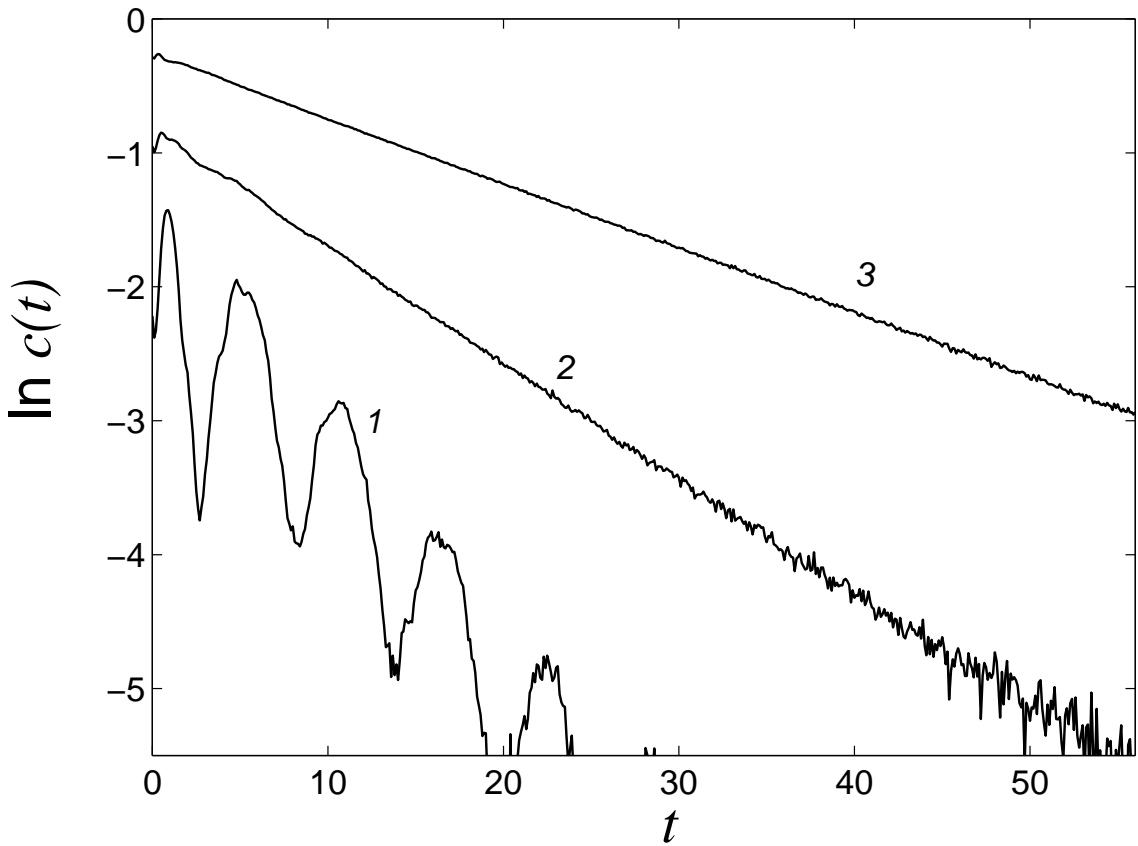}
\end{center}
\caption{\label{fig2}\protect\small
       Correlation function of the system of particles with $d=0.5$
       under temperatures $T=0.24$, 0.45 and 0.75 (curves 1, 2 and 3).
        }
\end{figure}

Numerical simulation of the dynamics demonstrates exponential decrease of the autocorrelation
$C(t)\sim \exp(-\alpha t)$ for all values of the diameter $0<d<2$ and temperature $T>0$ where
the simulation time is plausible from technical viewpoint. For low temperatures however the
exponential decrease is accompanied by oscillations with period corresponding to the frequency
of the vibrations near the potential minima (Fig. \ref{fig2}). The reason is that if the
temperatures are low, the concentration of transient particles decreases exponentially and
majority of the particles vibrates near the potential minima. It means that the 1D gas on the
on-site potential has finite heat conductivity. Coefficient of the exponential decrease of the
autocorrelation function
        \begin{equation}
        \alpha=-\lim_{t\rightarrow\infty}\frac{\ln C(t)}{t}
        \label{f15}
        \end{equation}
and coefficient of the heat conduction
        \begin{equation}
        \kappa=\int_0^\infty C(t)dt.
        \label{f16}
        \end{equation}
are computed numerically.

\begin{figure}[hp]
\begin{center}
\includegraphics[angle=0, width=0.7\textwidth]{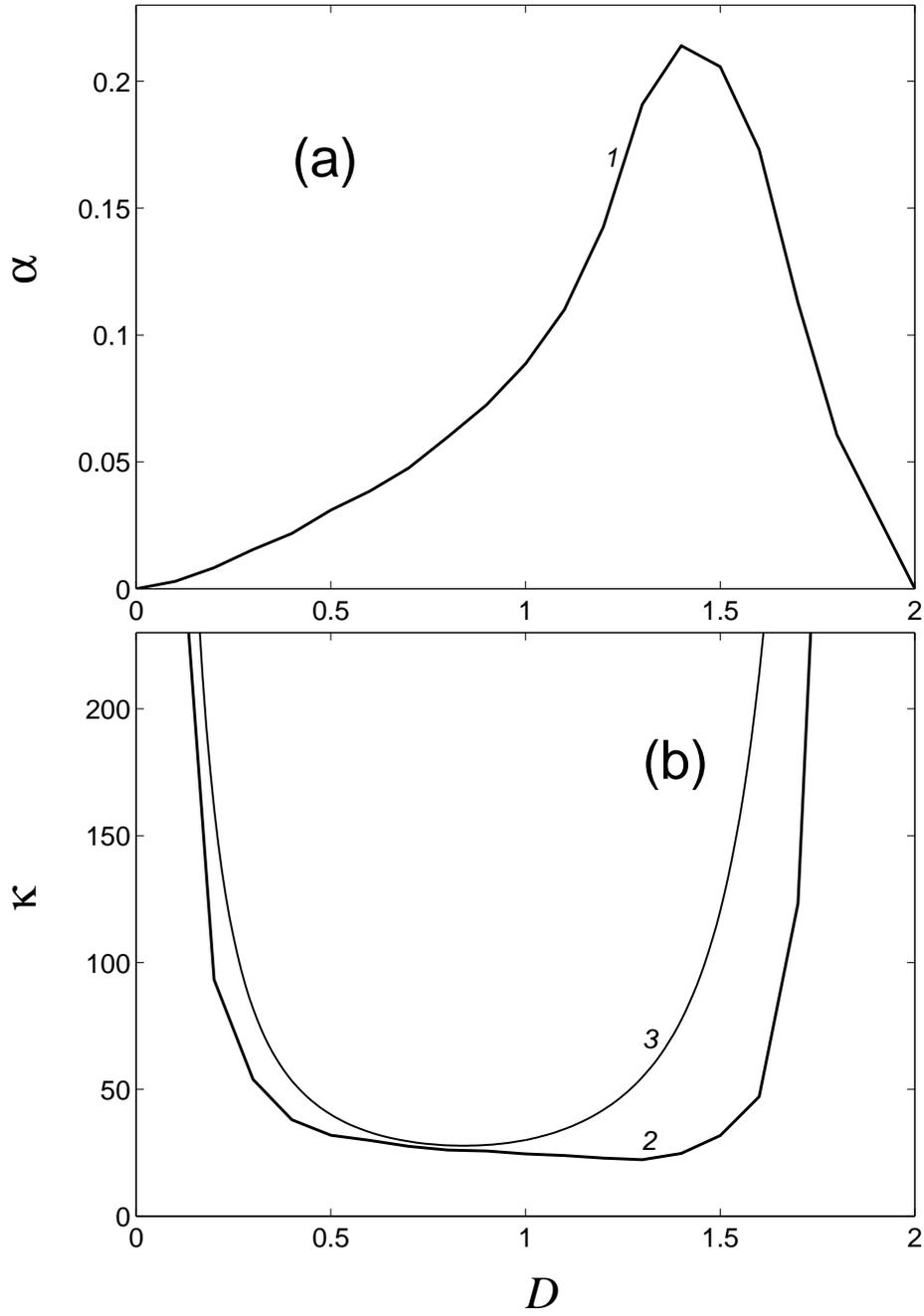}
\end{center}
\caption{\label{fig3}\protect\small
       Dependence of the coefficient of the exponential decrease of the autocorrelation
       function $\alpha$ (a) and the coefficient of the heat conduction $\kappa$ (b)
       on the particle diameter $d$ of 1D gas at T=1.
       Curves 1 and 2 correspond to piecewise linear on-site potential (\ref{f10}),
       curve 3 represent theoretical predictions according to formula (\ref{f11}).
        }
\end{figure}
\begin{figure}[hp]
\begin{center}
\includegraphics[angle=0, width=0.7\textwidth]{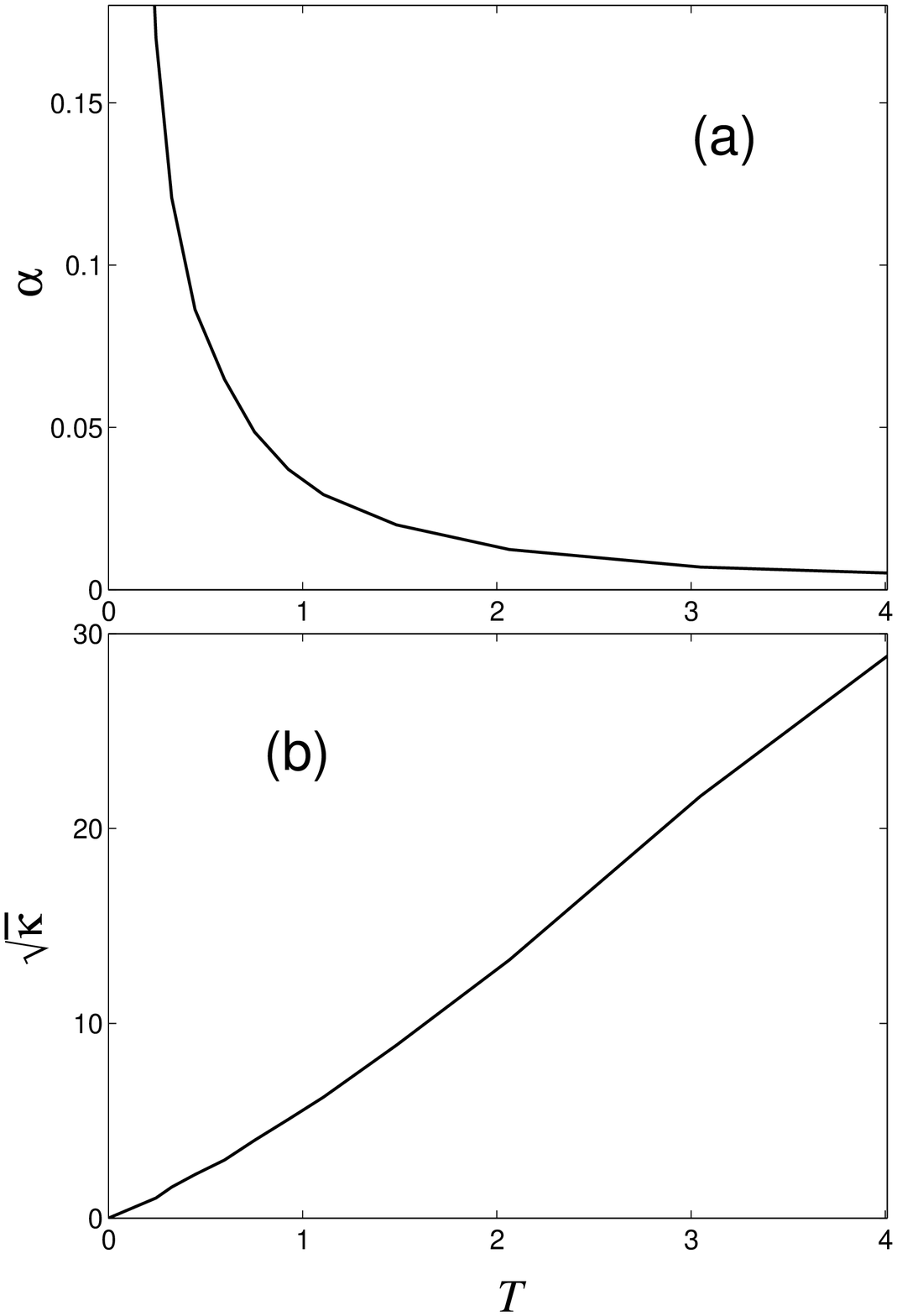}
\end{center}
\caption{\label{fig4}\protect\small
       Temperature dependence of exponent coefficient $\alpha$ (a) and heat
       conduction coefficient $\kappa$ (b) for particle diameter $d=0.5$.
        }
\end{figure}

Dependence of $\alpha$ and $\kappa$ on particle diameter $d$ is presented at Fig. \ref{fig3}.
Maximum of $\alpha$ and minimum of $\kappa$ is attained at $d=1.4$. As the temperature grows,
$\alpha$ decreases [Fig. \ref{fig4} (a)], and heat conduction $\kappa$ increases [Fig.
\ref{fig4} (a)].

Theoretical analysis of the heat conductivity presented above allows only approximate [although
rather reliable, see Figs. 3(b), 5] prediction of the numerical value of the heat conduction
coefficient $\kappa$. Still, the other question of interest is the asymptotic dependence of the
heat conduction on the parameters of the model. Formulae (\ref{f11}), (\ref{f12}) lead to the
following estimations:
        \begin{eqnarray}
        \kappa\sim T^{5/2}, &&~~\mbox{for}~~T\rightarrow \infty, \label{f17} \\
        \kappa\sim d^{-2}, &&~~\mbox{for}~~d\rightarrow +0, \label{f18} \\
        \kappa\sim (2-d)^{-3}, &&~~\mbox{for}~~d\rightarrow 2-0. \label{f19}
        \end{eqnarray}
These estimations should be compared to numerical results.

\begin{figure}[t]
\begin{center}
\includegraphics[angle=0, width=0.85\textwidth]{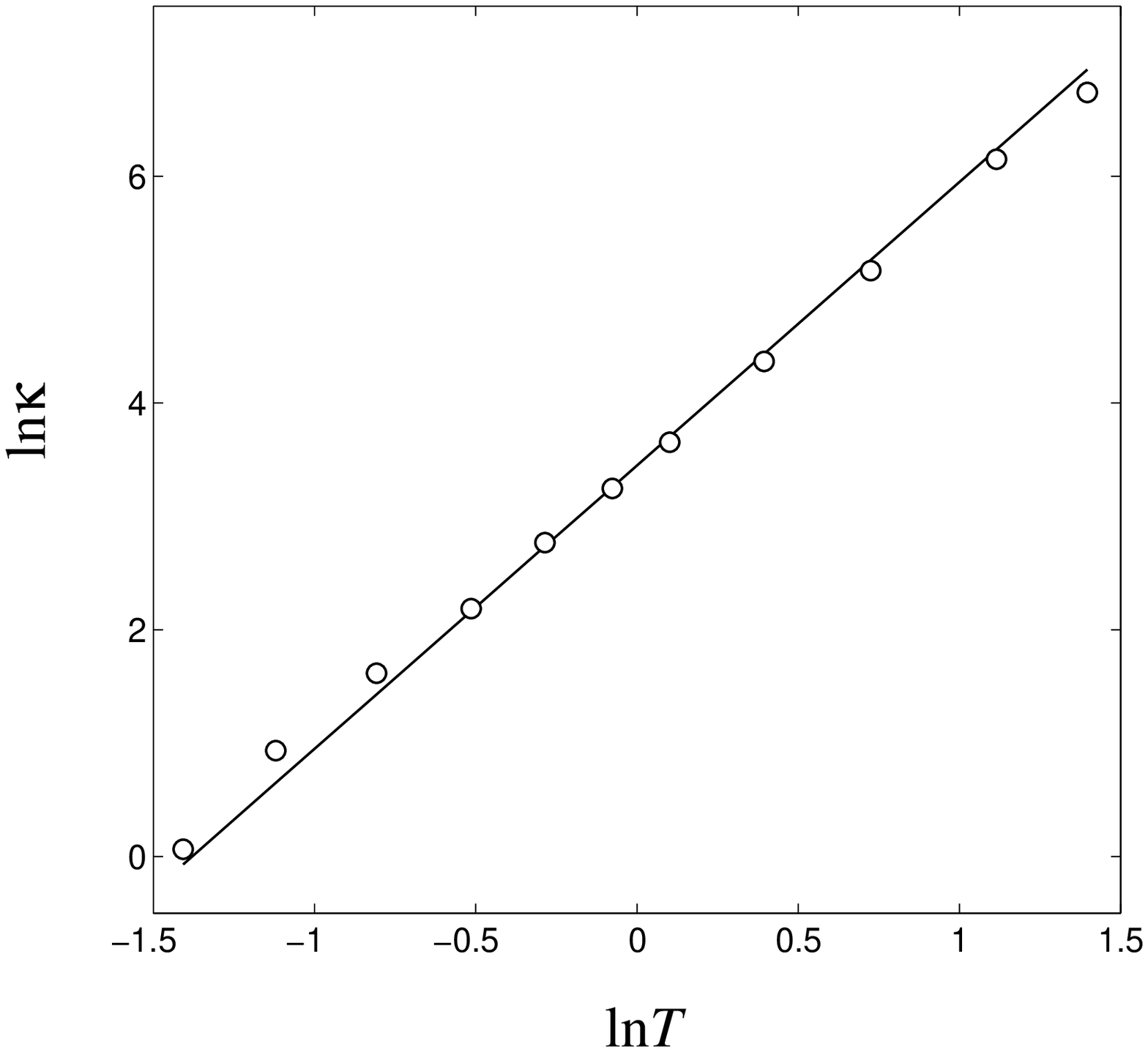}
\end{center}
\caption{\label{fig5}\protect\small
        Dependence of heat conduction coefficient on the temperature.
        The markers correspond to numerical results ($\ln\kappa$ versus $\ln T$),
        the straight line is  $\ln\kappa=2.5\ln T +3.45$, corresponding
        to (\ref{f17}). Particle diameter $d=0.5$ .
        }
\end{figure}
\begin{figure}[hp]
\begin{center}
\includegraphics[angle=0, width=0.75\textwidth]{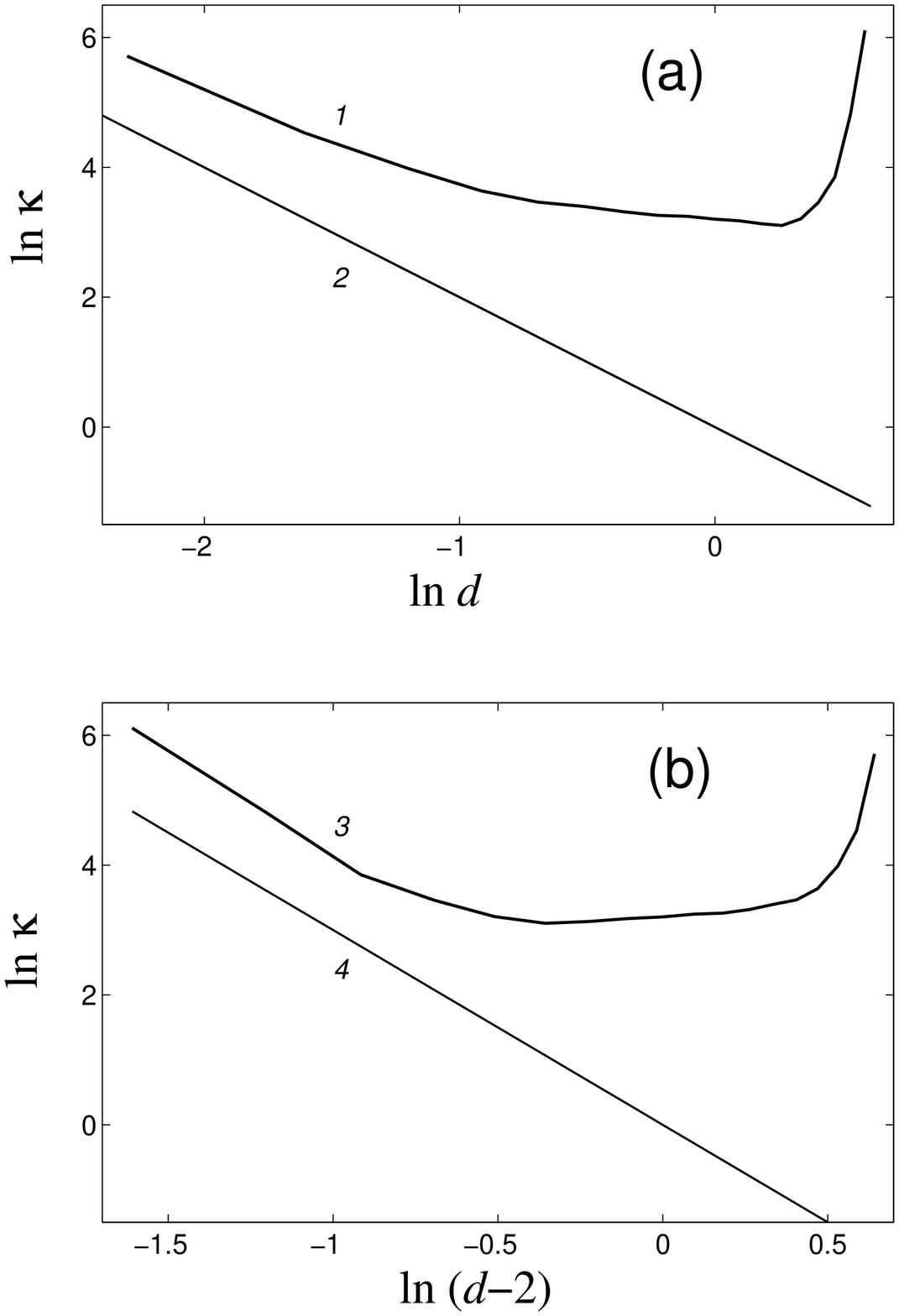}
\end{center}
\caption{\label{fig6}\protect\small
       Dependence of the heat conduction coefficient on the particle diameter
       (logarithmic coordinates, $\ln\kappa$ versus $\ln d$ (a) and versus $\ln (2-d)$ (b),
       curves 1 and 3). Lines $\ln\kappa=-2\ln d$ (curve 2) and $\ln\kappa=-3\ln (2-d)$
       (curve 4) correspond to relationships (\ref{f18}) and (\ref{f19}). Temperature $T=1$ .
        }
\end{figure}
\begin{figure}[hp]
\begin{center}
\includegraphics[angle=0, width=0.7\textwidth]{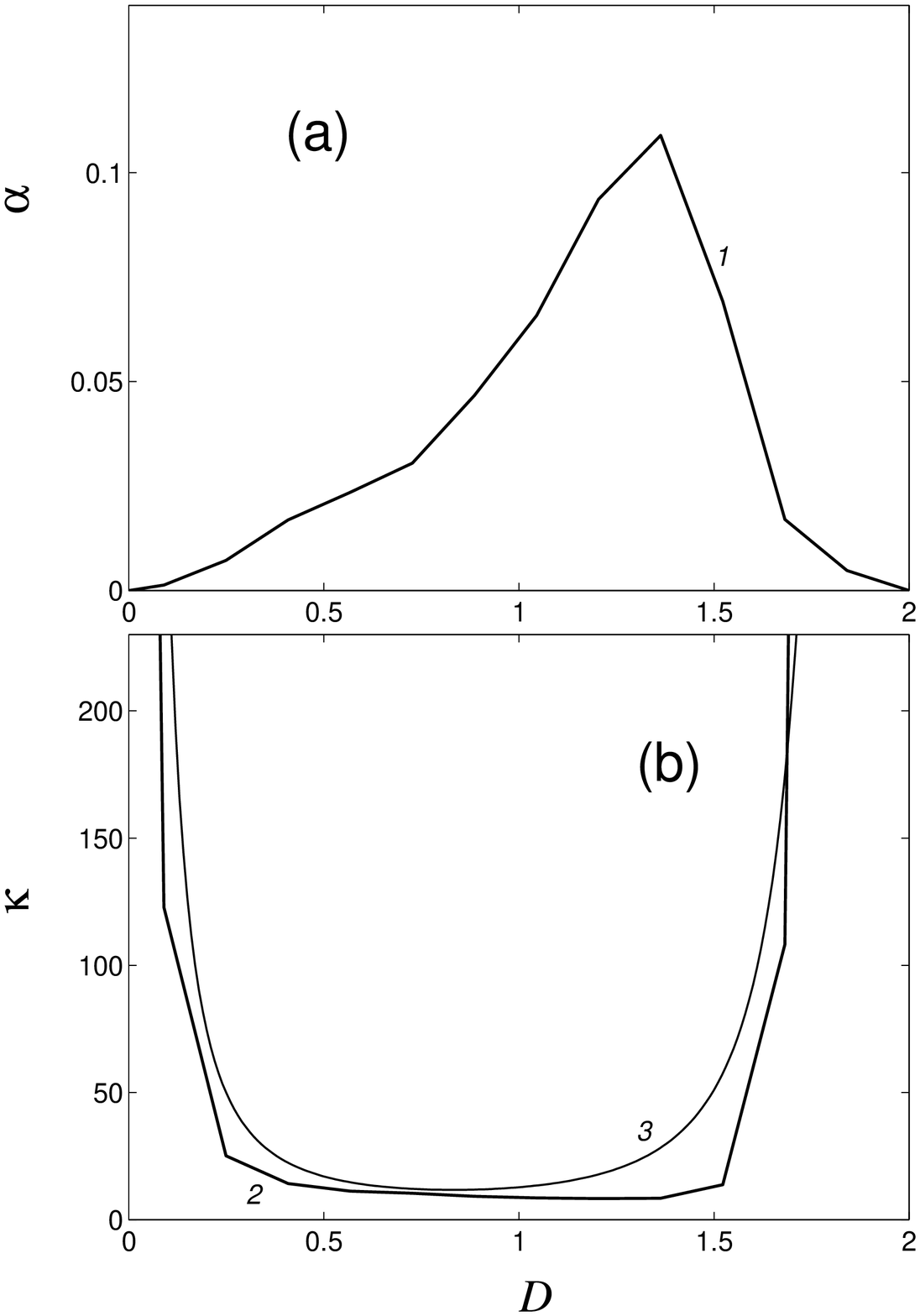}
\end{center}
\caption{\label{fig7}\protect\small
       Dependence of the coefficient of the exponential decrease of the autocorrelation
       function $\alpha$ (a) and the coefficient of the heat conduction $\kappa$ (b)
       on the particle diameter $d$ of 1D gas at T=1.
       Curves 1 and 2 correspond to smooth on-site potential (\ref{f20}),
       curve 3 represent theoretical predictions according to formula (\ref{f11}).
        }
\end{figure}

In order to check estimation (\ref{f17}) we consider the dependence of the logarithm of the
heat conduction $\ln\kappa$ on the logarithm of the temperature $\ln T$. From Fig. \ref{fig5}
it is clear that in accordance with (\ref{f17}) $\ln\kappa$ grows as $2.5\ln T$ as
$T\rightarrow \infty$. Fig. \ref{fig6} (a) demonstrates that as $d\rightarrow +0$, the
logarithm $\ln\kappa$ grows as $-2\ln d$, in accordance with (\ref{f18}). Fig. \ref{fig6} (b)
demonstrates that as $d\rightarrow 2-0$, the logarithm $\ln\kappa$ grows as $-3\ln(2-d)$, in
accordance with (\ref{f19}). So, it is possible to conclude that analytical estimations
(\ref{f17}), (\ref{f18}) and (\ref{f19}) fairly correspond to the numerical simulations data.

The above analytical estimations imply that the type of dependence of characteristic exponent
$\alpha$ and heat conductivity $\kappa$ on diameter $d$ and temperature $T$ does not depend on
the concrete shape of on-site potential $U(x)$ -- actually, only its finiteness and periodicity
do matter. Piecewise linear periodic potential (\ref{f10}) was chosen since it allowed
essential simplification of the numerical procedure. For comparison we have considered also the
smooth sinusoidal periodic potential
               \begin{equation}
               U(x)=[1-\cos(\pi u)]/2   \label{f20}
               \end{equation}
with period 2 and amplitude $U_0=1$, similarly to potential (\ref{f10}).

Potential (\ref{f20}) does not allow exact integration and requires standard numerical
procedures. Therefore it is convenient also to change rigid wall potential (\ref{f2}) by smooth
Lennard-Jones potential
              \begin{equation}
              V(\epsilon; r)=\epsilon\left(\frac{1}{r-d}-\frac{1}{2-d}\right)^2. \label{f21}
              \end{equation}
Parameter $\epsilon>0$ characterizes rigidity of the potential, the hard-particle potential
being the limit case:
              $$
              V(r)=\lim_{\epsilon\rightarrow +0} V(\epsilon; r).
              $$

Methods of computing of the autocorrelation function $C(t)$ and the heat conduction coefficient
$\kappa$ in 1D chain with analytic potentials of interaction are described in \cite{p25} in
detail. It should be mentioned that in order to get close to the limit of the hard particles we
should use small values of $\epsilon$ (at the temperature $T=1$ value $\epsilon=0.01$ was
used). It implies rather small value of the integration step. (We used standard Runge-Kutta
procedure of the fourth order with constant integration step $\Delta t=0.0001$). Therefore for
the case of hard (or nearly hard) particles the simulation with smooth on-site potential
(\ref{f20}) is far more time-consuming than the simulation with piecewise linear potential
(\ref{f2}).

In the case of hard particles with smooth on-site potential the autocorrelation function $C(t)$
decreases exponentially as $t\rightarrow\infty$ for all range $0<d<2$, $T>0$, i.e. the heat
conduction converges. Fig. \ref{fig7} demonstrates that the type of dependence of $\alpha$ and
$\kappa$ on parameters $d$ and $T$ is similar for piecewise linear potential (\ref{f2}) and
sinusoidal potential (\ref{f20}) (although numerical values $\alpha$ and $\kappa$ vary
slightly). For this potential function $F(d)=\frac12\sin^2(\pi d/2)$. It confirms that type of
heat conduction does not depend on concrete choice of on-site potential function.

\section{Conclusion}

We have considered the heat conduction process in the 1D lattice of hard particles with
periodic on-site potential. Analytical treatment predicts that for zero diameter of the
particles the system will be completely integrable regardless the exact shape of the on-site
potential. Therefore the heat conductivity will be infinite. For any nonzero size of the
particles the heat transfer is governed by diffusion of quasi\-particles giving rise to finite
heat conductivity. The value of the heat conduction coefficient computed by the analytical
treatment is in line with numerical simulation data. This coincidence is very profound if
speaking about the asymptotic scaling behavior of the heat conduction coefficient in the cases
of small and large particle sizes, as well as for the case of high temperatures. The
characteristic behavior of the heat conduction coefficient does not depend on the exact shape
of the on-site potential function.

The above results mean that there exists a new class of universality of 1D chain models with
respect to their heat conductivity. The limit case of zero-size particles is integrable, but
the slightest perturbation of this integrable case by introducing the nonzero size leads to
drastic change of the behavior -- it becomes diffusive and the heat conduction coefficient
converges. It should be stressed that this class of universality, unlikely the systems with
conserved momentum, cannot be revealed by sole numerical simulation. The reason is that the
correlation length (as well as the heat conduction coefficient) diverges as the system
approaches the integrable limit; therefore any finite capacity of the numerical installation
will be exceeded. That is why the analytical approach is also necessary.

The authors are grateful to Russian Foundation of Basic Research (grant 01-03-33122) and to RAS
Commission for Support of Young Scientists (6th competition, grant No. 123) for financial
support. O.V.G. is grateful to Russian Science Support Foundation for the financial support.

\end{document}